\documentclass[12pt]{article}

\newcommand{\be }{\begin{equation}}
\newcommand{\ee }{\end{equation}}

\def\hl{\hat{\lambda}}
\def\hs{\hat{\sigma}}
\def\hr{\hat{\rho}}

\def\H{{}^\star\!H}

\def\ch{{\cal{H}}}

\def\hm{\hat{\mu}}
\def\hn{\hat{\nu}}
\def\cc{\cal{C}}

\usepackage{amsmath}
\def\H{{}^\star\!H}
\def\c{\cal{C}}
\def\cd{\cal{D}}

\begin{document}
\vspace*{-.6in}
\thispagestyle{empty}
\begin{flushright}
USB-preprint: SB/F/288-01
\end{flushright}
\baselineskip = 20pt

\vspace{.5in}
{\Large
\begin{center}
{Master Canonical Action and BRST Charge of the M Theory Bosonic
Five Brane}
\end{center}}

\begin{center}
A. De Castro{\footnote{E-mail address: alex@fis.usb.ve}} and A.
Restuccia\footnote{E-mail address: arestu@usb.ve}$^{\dagger}$\\
\emph{Universidad Sim\'on Bol\'{\i}var, Departamento de
F\'{\i}sica, AP 89000, Caracas 1080-A, Venezuela\\
{$^{\dagger}$ Department of Mathematics, Kings College, London}}
\end{center}
\vspace{.5in}

\begin{center}
\textbf{Abstract}
\end{center}

\begin{quotation}
\noindent A complete analysis of the canonical structure for a gauge fixed PST
bosonic five brane action  is performed.
This canonical formulation is quadratic in the dependence on the
antisymmetric field and it has second class constraints. We remove
the second class constraints and  a master canonical action with
only first class constraints  is proposed. The nilpotent BRST
charge and its BRST invariant effective theory is constructed. The
construction does not assume the existence of the inverse of the
induced metric. Singular configurations are then physical ones. We
obtain the physical Hamiltonian of the theory  and analyze its stability
properties. Finally, by studying the algebra of
diffeomorphisms we find under mild assumptions the general
structure for the Hamiltonian constraint for theories invariant
under 6 dimensional diffeomorphisms and we give an algebraic
characterization of the constraint associated with the bosonic
five brane action. We also identify the constraint for the
bosonic five brane action upgraded with a cosmological term, it contains 
 a Born-Infeld type term.
\end{quotation}
\vfil

\newpage

\pagenumbering{arabic}

\section{Introduction}

In recent years the M5-brane in $D=11$ has acquired a protagonist
role in understanding the duality relations in the M-theory
framework. The degrees of freedom of 6d world volume theory of the
M5-brane are associated with the super-symmetric tensor multiplet
$N=(2,0)$. This multiplet contains a 2-form gauge field $B_{MN}$
with a self-dual field strength, five scalars and two chiral
spinors. The presence of this chiral gauge field was an obstacle
to the formulation of a covariant world volume M5-brane action. P.
Pasti, D. Sorokin and M.Tonin  in 1997 {\cite{Pasti1}}  found a
way of dealing with this obstacle  using the auxiliary scalar
field approach,  thus  allowing the formulation of a covariant
super-5-brane action with its usual $\kappa$-invariance
{\cite{Pasti2}}.  However, in order to obtain an operatorial or
functional integral quantum formulation of the theory, for example
to analyze the quantum stability properties of the theory, it
seems necessary  to eliminate by partial gauge fixing the
auxiliary scalar field of the PST approach, since it is present in
the denominator of the Lagrangian. The covariant field equations
for the super $M5$-brane were first obtained in {\cite{HS}},
{\cite{HSW}} and {\cite{BLW}} using the superembedding approach.

 An action for the M5-brane was independently
formulated in a non manifestly covariant way in {\cite{JS3}}, it
corresponds to a gauge fixed version of the covariant proposal.
There are two partial gauge fixing conditions  that naturally
eliminate off the auxiliary scalar field of the PST approach. One
of them corresponds to fixing the scalar field as the world volume
time and the other as a local world volume spatial coordinate.The
latest corresponds to the formulation in {\cite{JS3}}. In the
first case the action is first-order in time derivatives of the
antisymmetric field and in the second one is higher order in time
derivatives, moreover it is non polinomial in the dependence of the
time derivatives  of the antisymmetric field. The first case,
although has second class constraints, allows a direct canonical
analysis. We remark that the PST formulation as well as the
formulation in {\cite{JS3}} assume the existence of the inverse of
the induced metric.

The canonical Hamiltonian for the 5-brane was first obtained in
{\cite{unpublished}, {\cite{GGT}} and also discussed in
{\cite{townsend1}}, however in order to analyze the stability
properties of the 5-brane and to compare them with the $D=11$
supermembrane theory it is relevant to go one step further, to
include also in the analysis the configurations with zero
determinant of the induced metric since as we will show they are
physical configurations of the theory,  and study then the
physical Hamiltonian. That is, the Hamiltonian describing the
dynamics of the physical degrees of freedom once the gauge ones
have been eliminated by an admissible gauge fixing procedure.

When the theory is reduced to the subspace of solutions of the
field equations for which the induced metric is flat, the Lagrangian
of the 5-brane reduces to the one in {\cite{JS1}} whose  canonical
analysis was obtained in {\cite{Lamamiaqui}}. An interesting
feature of {\cite{Lamamiaqui}} was that the second class
constraints of the original formulation were eliminated giving
rise to a canonical Lagrangian with first class constraints only.
In this case, its double dimensional reduction yield a canonical
formulation allowing two covariant gauge fixing procedures, one of
them gives the Hamiltonian formulation of the 4-brane and the
other one its dual in terms of the antisymmetric field. The
analysis was performed even though the dependence on the time
derivatives of the antisymmetric field is non polinomial. The
resulting Hamiltonians contain in both cases the typical
Born-Infeld structure  for the field strength of the gauge vector
field and antisymmetric one.

In this work  we give first  a complete analysis of the canonical
structure of the bosonic sector of the 5-brane. We start from the
PST action and consider the partial gauge fixing where the scalar
field is fixed as the world volume time. The canonical
formulation of the $M5$-brane turns out to be quadratic in the
dependence on the antisymmetric field, and has second class
constraints together with the first class ones which generates the
symmetries of the theory. The Hamiltonian in {\cite{GGT}} reduces
to the one we obtain once their constraints are properly combined.

We remove the second class constraints preserving the locality of
the field theory and ending up with a master canonical action
which only contains first class constraints and is well defined
even for singular induced metrics. We refer as singular
configurations the ones for which the determinant of the spatial
part of the induced metric is zero at some point  or neighborhood
of the world volume. The  algebra of
the 6 dimensional diffeomorphisms generated by the first class
constraints is explicitly given. It is an open algebra.

This is the first step in the construction of an unconstrained
extended phase where a realization of the nilpotent BRST charge
may be possible. The BRST charge is a fundamental geometrical
object in the formulation of the quantum field theory associated
to the canonical Lagrangian. It may also be an important
geometrical object in the construction of the Seiberg-Witten  map
{\cite{Seiberg:1994rs}} that relates the 5-brane theory formulated
in terms of a commutative geometry to a formulation in terms of an
associated noncommutative one. The Seiberg-Witten map is a one to
one correspondence between gauge equivalent classes, which
correspond to BRST equivalence classes in the extended phase
space. In particular, this implies that the cohomology classes of
the BRST operators in the commutative realization and in the non
commutative one are in one to one correspondence. We construct the
nilpotent BRST charge of the theory and its BRST invariant
effective theory.  The construction involves several steps beyond
the standard construction for a closed algebra. In fact, one has
to introduce the higher order structure functions of the open
algebra.  We obtain its physical Hamiltonian and analyze its
stability properties remarking that although the Hamiltonian is
quadratic on the dependence on the antisymmetric field, its
reduction to a flat induced metric gives the Born-Infeld type of
Hamiltonian.
 The improvement is important when the operatorial
formulation of the BRST invariant effective action is considered.

Finally, by studying the algebra of diffeomorphisms  we find the
most general structure for the Hamiltonian constraint in terms of
the membrane  maps and the antisymmetric field. The corresponding
canonical Hamiltonian describe field theories which are invariant
under diffeomorphisms over a 6 dimensional world volume with a
chiral gauge field. The unique Hamiltonian constraint with
quadratic dependence on the antisymmetric field corresponds to the
M5-brane. We also identify the constraint associated to the
bosonic M5-brane action upgraded with a cosmological term,it
involves a Born-Infeld type of term. From the algebraic point of
view, the Hamiltonian constraint  for the M5-brane is
characterized by being the unique one which is polinomial in its
dependence on the field strength of the antisymmetric field and is
well defined even for singular induced metrics. The
BRST invariant action, that  we will construct, is characterized also by
the same property. Its Lagrangian density is well behaved even for
singular configurations of the induced metric. Hence, singular
configurations of the metric are allowed as physical
configurations. In distinction, the PST covariant action, as well
as the one in {\cite{JS3}}, assume the existence of the inverse of
the induced metric at any point of the world volume. In the case
of the D=11 Supermembrane, the singular configurations are also
physical configurations. This has important consequences with
respect to the quantum properties of the theory.  In fact, the
existence of these configurations together with the supersymmetry
render the spectrum continuous from zero to infinite. They are
also related to the topology changes on the embedded surface in
the target space, without changing the energy  of the
supermembrane. It is thus very important to take them into account
in any analysis of the theory.

Another consequence of the existence of singular configurations is
that the static ``gauge'' is not a correct gauge
fixing of the theory, since giving a singular configuration one
cannot gauge transform it to a non singular one (in the static
gauge the spatial part of the induced metric is always non
singular). If one restricts the theory to the space of nontrivial
higher order bundles {\cite{mariohob}} or equivalent to non
trivial wrapping of the 5-brane on the target space, then the
singular configurations are avoided and it becomes a correct gauge
fixing condition. However the quantum stability analysis of the
theory must include all canonical admissible configurations and it
is in that case that the static gauge is not allowed. In
distinction the light cone gauge  is always an admissible gauge.
In fact, even the singular configurations can be gauged to it, and
of course they remain to be singular.


\section{Canonical analysis of the bosonic 5-brane action}\label{CA}

We start from the PST action for the M5-brane and consider the
gauge in which the scalar field is proportional to the world
volume time. This gauge fixing, associated to the gauge symmetry of the
auxiliar scalar field, may be implemented  directly into the action, since the
Fadeev-Popov procedure gives a constant contribution to the measure of the
functional integral.   We obtain the following Lagrangian density

\begin{equation}\label{LS}
L=2\sqrt{-\det
\left(G_{MN}+G_{M\rho}G_{N\lambda}\H^{\rho\lambda}\sqrt{-G_0}\right)}+\frac{1}{2}
\H^{\mu\nu}\partial_0 B_{\mu\nu}+\frac{1}{4}
\epsilon_{\mu\nu\rho\lambda\sigma}
\frac{G^{0\rho}}{G^{00}}\H^{\mu\nu}\H^{\lambda\sigma}
\end{equation}

Where $B$ denotes  the antisymmetric gauge field and $H$ is the
self-dual field strength $H=dB$. The 6 dimensional  world volume
indices are denoted by $M,N=0,1,\cdots,5$ while  the spatial ones
by $\mu\nu=1,\cdots,5$. $G_{MN}$ in terms of the 5-brane maps
$X^a$ is given by the induced expression  $G_{MN}=\partial_M
X^a\partial_N X_a$, where $a$ denotes the $D=11$ Minkowski
indices. In the spatial 5-dimensional world volume we denote:

\begin{equation}
H_{\rho\lambda\sigma}=\partial_\rho B_{\lambda\sigma}+\partial_\sigma
B_{\rho\lambda}+\partial_\lambda B_{\sigma\rho}
\end{equation}

\begin{equation}
\H^{\mu\nu}\equiv \frac{1}{6}\epsilon^{\mu\nu\gamma\delta\lambda}H_{\gamma\delta\lambda}
\end{equation}

Where ${\H}^{\mu\nu}$ is a contravariant density.  In order to
analyze the Hamiltonian structure of this action it is convenient
to introduce the {\bf{ADM}} parametrization of the metric

\begin{eqnarray}G_{00}&\equiv&-(n^2-N_\lambda
N^\lambda)=\dot{X}^a\dot{X}_a\cr
G_{0\mu}&\equiv&N_\mu=\dot{X}^a\partial_{\mu}X_a\cr G_{\mu\nu}
&\equiv&g_{\mu\nu}=\partial_\mu X^a\partial_\nu X_a\cr
N^{\mu}&\equiv&g^{\mu\nu}N_\nu,\cr G^{00}&=&-(\frac{1}{n^2}),\cr G^{0\mu}&=&\frac{N^\mu}{n^2},\cr
G^{\mu\nu}&=&g^{\mu\nu}-\frac{N^\mu N^\nu}{n^2},\cr
\det(G_{MN})&=&-n^2\det(g_{\mu\nu}).
\end{eqnarray}

They allow us to  rewrite (\ref{LS}) in the form

\begin{equation}
L=2n\sqrt{gM}-\frac{1}{4}N^\rho
\hat{V}_\rho+\frac{1}{2}\H^{\mu\nu}\partial_0 B_{\mu\nu}
\end{equation}

where

\begin{eqnarray}
g&=&\det{g_{\mu\nu}}\cr {M}&\equiv&1+\hat{y}+\hat{z}\cr
\hat{y}&\equiv&\frac{1}{2}g^{-1}\H^{\alpha\beta}\H_{\alpha\beta}\cr
\hat{z}&\equiv&\frac{1}{64}g^{-1}g^{\mu\nu}\hat{V}_\mu\hat{V}_\nu\cr
\hat{V}_\mu&=&\epsilon_{\mu\alpha\beta\gamma\delta}\H^{\alpha\beta}\H^{\gamma\delta}
\end{eqnarray}

The spatial world volume indices are raised and lowered with the
induced metric, the need of using the inverse of $g_{\mu\nu}$ will
be later on relaxed. The term  $g M$ is precisely

\begin{equation}
\det(g_{\mu\nu}+g^{-1/2}\H_{\mu\nu})=gM
\end{equation}

The conjugate momenta to $X^a$ may be directly evaluated. It is

\begin{equation}\label{pi1}
\Pi_a=2\frac{g^{1/2}}{n}(-\dot{X}_a+N^{\lambda}\partial_\lambda
X_a)M^{1/2} -\frac{1}{4}\hat{V}^\rho\partial_\rho X_a
\end{equation}

We can guess that the diffeomorphisms constraints have a similar
diffeomorphisms constraints structure as in string and membrane
theory. We finally obtain

\begin{eqnarray}
\label{vinc1}\hat{\phi}&=&\frac{1}{2}\Pi^2+g(2\hat{y}+2)=0\\
\label{vinc2}\hat{\phi}_\alpha&=&\Pi_a\partial_\alpha
X^a+\frac{1}{4}\hat{V}_\alpha=0
\end{eqnarray}

The conjugate momenta to $B_{\mu\nu}$ will be denoted $P^{\mu\nu}$ and
satisfies the constraint

\begin{equation}\label{vinc3}
\Omega^{\mu\nu}=P^{\mu\nu}-\H^{\mu\nu}=0
\end{equation}

The standard canonical analysis shows that (\ref{vinc1}),
(\ref{vinc2}) and (\ref{vinc3}) are the only constraints of the
theory and they are a mixture of first and second class
constraints.  The canonical Hamiltonian is a linear combination of
these constraints

\begin{equation}\label{Hcan}
{\ch}_c=\Lambda\hat{\phi}+\Lambda^{\alpha}\hat{\phi}_{\alpha}+\Lambda_{\mu\nu}\Omega^{\mu\nu}
\end{equation}

We will now consider a more general canonical formulation with
first class constraints only, which under partial gauge fixing
reduces to (\ref{Hcan}). Such a formulation can be obtained by
several approaches. Following  {\cite{Restuccia1}},  we deduce
from (\ref{vinc3})the first class constraints for the
antisymmetric field

\begin{eqnarray}
\Omega^{5i}&=&P^{5i}-\H^{5i}=0\label{fclass1}\\
\Omega^{j}&=&\partial_{\mu}P^{\mu j}=0\label{fclass2},\quad
i=1,2,3,4.
\end{eqnarray}

The diffeomorphisms constraints must be modified in order to
obtain a complete set of first class constraints and we propose
the following structure:

\begin{eqnarray}
\phi&=&\frac{1}{2}\Pi^2+g(2y+2)=0\label{fclass3}\\
\phi_\alpha&=&\Pi_a\partial_\alpha
X^a+\frac{1}{4}V_\alpha=0\label{fclass4}
\end{eqnarray}

where

\begin{eqnarray}
y&=&\frac{1}{8}g^{-1}(P^{\alpha\beta}+\H^{\alpha\beta})(P_{\alpha\beta}+\H_{\alpha\beta})\
\label{y}\\
V_\mu&=&\frac{1}{4}\epsilon_{\mu\alpha\beta\gamma\delta}(P^{\alpha\beta}+\H^{\alpha\beta})
(P^{\gamma\delta}+\H^{\gamma\delta})\label{V}
\end{eqnarray}

 $\Omega^{5i}$ and
$\Omega^{i}$    commute between themselves and with $\phi$ and
$\phi_\alpha$.  The simplectic space of the antisymmetric field
and its conjugate momenta has been decomposed in terms of the
local commuting coordinates

\[
P^{\mu\nu}+\H^{\mu\nu}
\]

and

\[
P^{\mu\nu}-\H^{\mu\nu}
\]

with Poisson brackets

\begin{equation}
\{P^{\mu\nu}+\H^{\mu\nu},P^{\alpha\beta}-\H^{\alpha\beta}\}=0
\end{equation}

the latest coordinate may be fixed by a gauge condition associated
to constraints (\ref{fclass1}) and (\ref{fclass2}). The procedure
to obtain (\ref{fclass1}),  (\ref{fclass2}), (\ref{fclass3}) and
(\ref{fclass4}) is deductive, however  we have only presented the
resulting expressions whose required properties may be directly
checked {\cite{Restuccia1}}. $\phi$ and $\phi_\alpha$ generates
the algebra of diffeomorphism on the world volume: The canonical
Hamiltonian is then the linear combination of the first class
constraints. We have broken even further the manifest covariance
of the formulation, nevertheless we have now a polinomic
formulation of the theory in terms of first class constraints
only. We remark that the square root characteristic of the
Born-Infeld action for the 5-brane has now disappear. These
properties ensure the construction of a polinomic physical
Hamiltonian and a polinomic BRST invariant formulation of the
theory.  We will consider both points in the following sections.

The algebra generated by the first class constraints
(\ref{fclass1}),  (\ref{fclass2}), (\ref{fclass3}) and
(\ref{fclass4}) is given by:

\begin{eqnarray}
\{\phi_\rho,\phi^{\prime}_\mu\}&=&\phi_\mu\partial_\rho\delta+\phi^{\prime}_\rho\partial_\mu\delta
-(\partial_\gamma{P}^{\gamma\delta})l^{\alpha\beta}\epsilon_{\mu\delta\rho\alpha\beta}\cdot\delta\\
\{\phi_\rho,\phi^{\prime}\}&=&(\phi+\phi^{\prime})\partial_\rho\delta-
4(\partial_\lambda{P}^{\lambda\alpha})l_{\alpha\rho}\cdot\delta\\
\{\phi,\phi^{\prime}\}&=&({\cc}^{\rho\sigma}\phi_\rho+
{\cc}^{\prime\rho\sigma}\phi_\rho^{\prime})\partial_\sigma\delta
\end{eqnarray}

where

\begin{equation}
{\cc}^{\sigma\lambda}=4(gg^{\sigma\lambda}+g_{\alpha\beta}l^{\alpha\sigma}l^{\beta\lambda}),
\;\;l^{\alpha\beta}\equiv\frac{1}{2}(P^{\alpha\beta}+\H^{\alpha\beta})
\end{equation}

We remark that the first class constraints, as well as the
structure functions of the algebra, depend only on $g_{\mu\nu}$
and not on its inverse. This is an important difference with
respect to the covariant formulation of the theory, where the
metric is assumed to have an inverse. It is well known in the case
of the $D=11$ supermembrane that the singular configurations,
which in this theory are string like configurations, are
responsible for the continuous spectrum of the Hamiltonian as well
as for topological changes of the brane. {\cite{deWit:1988ig}},  {\cite{dWLN}} 
{\cite{nicolai}}.

\section{BRST effective action}

We have succeeded in finding a formulation of the 5-brane which is
polinomic on the fields and with first class constraints only. We
will now construct the associated nilpotent BRST charge and the
BRST invariant effective action. The BRST charge is the
fundamental geometrical object in the quantum analysis of the
theory and in the construction of the map relating the 5-brane
formulation to a non commutative geometry.

The effective action will depend on the covariant induced metric
only, and not  in its inverse, so is well behaved even on singular
configurations  where the determinant of the induced metric
becomes zero.

In order to have a realization of the generators of spatial
diffeomorphisms  $\phi_\rho$, in the whole space of geometrical
objects including the antisymmetric field and its conjugate
momenta we will consider a reduction of the phase space by
restricting to configurations satisfying

\begin{equation}\label{divP}
\partial_\mu{P}^{\mu\nu}=0
\end{equation}

We may performed this restriction starting with the whole extended
phase space and imposing finally at the level  of  the effective
action  a  canonical gauge fixing  condition associated to
(\ref{divP}):

\begin{equation}\label{gauge}
\chi(B,P)=0
\end{equation}
depending only on $B$ and $P$ fields.

Otherwise we may start restricting the phase space by conditions
(\ref{divP}) and (\ref{gauge}) and working out the nilpotent BRST
charge on the subspace of phase space. In our case, the latest
approach becomes slightly more direct. This is so because
$\partial_\mu{P}^{\mu\nu}$ commutes with all constraints of the
theory and with all phase space coordinates except $B_{\mu\nu}$.
However $B_{\mu\nu}$ appears in the constraints only through
$\H^{\alpha\beta}$, which commutes with
$\partial_\mu{P}^{\mu\nu}$. Consequently, in the  BRST
construction we may work  directly with Poisson brackets even so
we are restricted by (\ref{divP}) and (\ref{gauge}) (Dirac
brackets are the same as Poisson brackets). We will follow this
latest approach, for simplicity we require

\begin{equation}\label{gaugecorchete}
\{\chi,\chi^{\prime}\}=0
\end{equation}

The algebra of diffeomorphism obtained in section \S \ref{CA} is
an open algebra, that is, the structure functions depend on the
fields. Consequently, the construction of the nilpotent BRST
charges requires several additional steps beyond the standard
construction for a closed algebra{\cite{H}}. They involve the
higher order structure functions of the algebra. We introduce the
ghosts $C$ and $C^\rho$ associated to $\phi$ and $\phi_\rho$and
its conjugate momenta $\mu$ and $\mu_\rho$ respectively. We start
from the Poisson brackets

\begin{equation}
\{\phi,\phi^{\prime}\}=(\cc^{\sigma\lambda}\phi_\sigma+
\cc^{\prime\sigma\lambda}\phi^{\prime_\sigma})\partial_\lambda\delta
\end{equation}

where

\begin{equation}
{\cc}^{\sigma\lambda}=4(gg^{\sigma\lambda}+g_{\alpha\beta}l^{\alpha\sigma}l^{\beta\lambda})
\end{equation}

It is convenient to extend the other generator of diffeomorphisms
$\phi_\rho$ from the beginning in order to simplify the
construction. We define

\begin{equation}
\widetilde{\phi}_\rho=\phi_\rho+2\mu\partial_\rho{C}+\partial_\rho\mu\cdot
C+\mu_\lambda\partial_\rho{C}^\lambda+\partial_\lambda(\mu_\rho{C}^\lambda)
\end{equation}

We then have

\begin{eqnarray}
\{\langle \xi^\rho\widetilde{\phi}_\rho \rangle,
C\}&=&-2\xi^\rho\partial_\rho{C}+
\partial_\rho(\xi^\rho{C})=-\xi^\rho\partial_\rho{C}+\partial_\rho\xi^\rho{C},\\
\{\langle \xi^\rho\widetilde{\phi}_\rho \rangle,
\mu\}&=&-\xi^\rho\partial_\rho\mu-2\partial_\rho\xi^\rho\mu
\end{eqnarray}

with the right density weights  to obtain

\begin{equation}
\{\langle \xi^\rho\widetilde{\phi}_\rho \rangle,
\mu{C}\}=-\xi^\rho\partial_\rho(\mu{C})-\partial_\rho\xi^\rho(\mu{C}),
\end{equation}

Consequently, we have

\begin{equation}
\{\langle \xi^\rho\widetilde{\phi}_\rho \rangle,
\langle{C}\phi\rangle\}=\langle-\partial_\rho(\xi^{\rho}C\phi)\rangle=0
\end{equation}

We will drops all boundary terms, that is we will assume a closed
compact spatial world volume. We also get

\begin{equation}
\{\langle \xi^\rho\widetilde{\phi}_\rho \rangle,
{C}^\mu\}=-\xi^\rho\partial_\rho{C}^\mu+\partial_\rho\xi^\mu{C}^\rho,
\end{equation}

since $C^\mu$ transforms as a contravariant vector

The introduction of $\widetilde{\phi}_\rho$ simplifies the
construction since its action on any new term in the BRST charge
reduces only to  count weights for objects which transform under
diffeomorphisms as densities.

We may start considering

\begin{equation}
Q=\langle
C\phi+C^\rho\widetilde{\phi}_\rho-\mu_\sigma{C}\partial_\lambda{C}{\cc}^{\sigma\lambda}
-C^\rho\mu_\lambda\partial_\rho{C}^\lambda \rangle + Q^H+Q^A\equiv
Q^C+Q^H+Q^A
\end{equation}

Where $Q^H$ denotes the terms involving the higher order structure
functions of the diffeomorphism  algebra. $Q^A$ denotes the
contributions to $Q$ of the constraints associated to the
antisymmetric field. It commutes with $Q^C+Q^H$, hence we will
consider it after the complete evaluation of $Q^C$ and $Q^H$. We
will systematically add the higher order terms in derivatives of
$C$ in order to ensure the nilpotency of $Q$.

We get

\begin{equation}
\{\langle{C}\phi\rangle,\langle {C}\phi \rangle\}=\langle
2C\partial_\lambda{C}{\cc}^{\rho\lambda}\phi_\rho \rangle
\end{equation}

\begin{equation}
2\{\langle{C}^\rho\widetilde{\phi}_\rho\rangle,
\langle-\mu_\sigma{C}\partial_\lambda{C}{\cc}^{\sigma\lambda}
\rangle\}=-\langle
2C\partial_\lambda{C}{\cc}^{\rho\lambda}\phi_\rho \rangle -
2\langle[\mu_{\hat{\lambda}}\partial_\rho{C}^{\hat{\lambda}}+\partial_{\hat{\lambda}}
(\mu_\rho{C}^{\hat{\lambda}})]C\partial_\lambda{C}{\cc}^{\rho\lambda}\rangle
\end{equation}

\begin{equation}
2\{\langle -\mu_\sigma{C}\partial_\lambda{C}{\cc}^{\sigma\lambda}
\rangle, \langle
-C^\rho\mu_{\hat{\lambda}}\partial_\rho{C}^{\hat{\lambda}}
\rangle\}=2\langle
C\partial_\lambda{C}{\cc}^{\sigma\lambda}\mu_\nu\partial_\sigma{C}^\nu
\rangle-2\langle{C}\partial_\lambda{C}{\cc}^{\sigma\lambda}\partial_{\rho}
({C}^{\rho}\mu_\sigma)\rangle
\end{equation}

\begin{equation}
2\{\langle{C}^\rho\widetilde{\phi}_\rho\rangle,
\langle{C}^\lambda\widetilde{\phi}_\lambda
\rangle\}=-2C^\nu\partial_\nu{C}^\rho\widetilde{\phi}_\rho
\end{equation}

\begin{equation}
2\{\langle{C}^\rho\widetilde{\phi}_\rho\rangle,
\langle-{C}^\nu\mu_\lambda\partial_\nu{C}^\lambda\rangle\}=
2C^\nu\partial_\nu{C}^\rho\widetilde{\phi}_\rho
\end{equation}

\begin{equation}
2\{\langle
{C}^{\hat{\nu}}\mu_{\hat{\lambda}}\partial_{\hat{\nu}}{C}^{\hat{\lambda}}
\rangle,\langle{C}^\nu\mu_\lambda\partial_\nu{C}^\lambda
\rangle\}=0
\end{equation}

All these brackets cancel between each other, which is the
standard situation for a closed  algebra. However, since
${\cc}^{\sigma\lambda}$ are field dependent, we have additional
contributions to the evaluation of the $\{Q,Q\}$.

We obtain

\begin{equation}\label{1termino}
2\{\langle{C}\phi\rangle,\langle
-\mu_\sigma{C}\partial_\lambda{C}{\cc}^{\sigma\lambda}
\rangle\}=\langle
2{\cc}_1^{\mu\nu,\sigma\lambda}\mu_\sigma{C}\partial_\lambda{C}\partial_\mu{C}\phi_\nu
\rangle
\end{equation}

where

\begin{equation}
{\cc}_1^{\mu\nu,\sigma\lambda}=4(-2gg^{\sigma\lambda}g^{\mu\nu}+gg^{\sigma\mu}g^{\nu\lambda}+
gg^{\sigma\nu}g^{\mu\lambda}-l^{\mu\sigma}l^{\nu\lambda}-l^{\nu\sigma}l^{\mu\lambda})
\end{equation}

The first contribution to $Q^H$ in then

\begin{equation}
Q^H=\langle
-\frac{1}{2}{\cc}_1^{\mu\nu,\sigma\lambda}\mu_\nu\mu_\sigma{C}
\partial_\lambda{C}\partial_\mu{C}+\cdots \rangle
\end{equation}

Notice that the commutator

\begin{equation}
\{\langle-C^\rho\mu_{\beta}\partial_\rho{C}^{\beta}\rangle
,-\frac{1}{2}{\cc}_1^{\mu\nu,\sigma\lambda}\mu_\nu\mu_\sigma{C}
\partial_\lambda{C}\partial_\mu{C}\},
\end{equation}

gives a contribution which combines with (\ref{1termino}) to give

\begin{equation}
\langle
2{\cc}_1^{\mu\nu,\sigma\lambda}\mu_\sigma{C}\partial_\lambda{C}\partial_\mu{C}
\widetilde{\phi}_\nu\rangle
\end{equation}

which is canceled by the commutator

\begin{equation}
2\{\langle{C}^\rho\widetilde{\phi}_\rho\rangle
,-\frac{1}{2}{\cc}_1^{\mu\nu,\sigma\lambda}\mu_\nu\mu_\sigma
\partial_\lambda{C}\partial_\mu{C}{C}\},
\end{equation}

${\cc}_1^{\mu\nu,\sigma\lambda}$ is again field dependent so there
are further contributions from the commutator  with
$\langle{C}\phi\rangle$.

The next contribution is from the commutators

\begin{equation}\label{42}
2\{\langle{C}{\phi}\rangle
,\langle-\frac{1}{2}{\cc}_1^{\mu\nu,\sigma\lambda}\mu_\nu\mu_\sigma
\partial_\lambda{C}\partial_\mu{C}{C}\rangle\} +
\{\langle-\mu_\sigma{C}\partial_\lambda{C}{\cc}^{\sigma\lambda}\rangle,
\langle -\mu_\sigma{C}\partial_\lambda{C}{\cc}^{\sigma\lambda}    \rangle\}
\end{equation}

which require the addition of a term

\begin{equation}\label{3}
\langle 8g(\mu^\lambda\partial_\lambda{C})^3C\rangle
\end{equation}

to $Q$.

The commutator of  $\langle C^\rho\widetilde{\phi}_\rho +
\mu_\lambda{C}^\rho\partial_\rho{C}^\lambda\rangle $  with
(\ref{3})

cancels the commutators in (\ref{42}).  Furthermore, the following
commutators

\begin{equation}\label{44}
2\{\langle{C}{\phi}\rangle ,\langle
8g(\mu^\lambda\partial_\lambda{C})^3C\rangle\} +
2\{\langle-\mu_\sigma{C}\partial_\lambda{C}{\cc}^{\sigma\lambda}\rangle,
\langle
-\frac{1}{2}{\cc}_1^{\mu\nu,\sigma\lambda}\mu_\nu\mu_\sigma{C}\partial_\lambda{C}
\partial_\mu{C} \rangle\}
\end{equation}

require a higher order term in $Q$. It is

\begin{equation}\label{4}
\langle 10g(\mu^\lambda\partial_\lambda{C})^4{C}\rangle
\end{equation}

The commutator of $\langle
C^\rho\widetilde{\phi}_\rho+\mu_\lambda{C}^\rho\partial_\rho{C}^\lambda\rangle$
with (\ref{4}) cancels the commutators in (\ref{44}).

Finally the commutators

\begin{equation}\label{46}
2\{\langle C\phi\rangle, \langle
10g(\mu^\lambda\partial_\lambda{C})^4{C} \rangle\}+\{ \langle
-\frac{1}{2}{\cc}_1^{\mu\nu,\sigma\lambda}
\mu_\nu\mu_\sigma{C}\partial_\lambda{C}\partial_\mu{C} \rangle,
\langle  -\frac{1}{2}{\cc}_1^{\mu\nu,\sigma\lambda}
\mu_\nu\mu_\sigma{C}\partial_\lambda{C}\partial_\mu{C}  \rangle\}
\end{equation}

require a final new terms in $Q$:

\begin{equation}\label{5}
\{\langle 12g(\mu^\lambda\partial_\lambda{C})^5{C}\rangle\}
\end{equation}

The commutator of $\langle
C^\rho\widetilde{\phi}_\rho\partial_\rho{C}^\lambda \rangle$ with
(\ref{5}) cancels the commutators in (\ref{46}).  Finally the
commutator of $\langle {C}\phi\rangle$ with (\ref{5}) gives zero
since the order in derivatives of $C$ is $6$. The nilpotency of
$Q$ has then been obtained.

The final form of the BRST charge is thus given by

\begin{eqnarray}\label{BRSTCH}
Q&=&\langle{C}\phi + C^\rho\widetilde{\phi}_\rho
+{\cc}^{\sigma\lambda}\mu_\sigma\partial_\lambda{C}{C}+\mu_\lambda{C}^\rho\partial_\rho{C}^\lambda\cr\cr
&-&
\frac{1}{2}{\cc}_1^{\mu\nu,\sigma\lambda}\mu_\sigma\partial_\lambda{C}\mu_\nu\partial_\mu{C}{C}+
8g(\mu^\lambda\partial_\lambda{C})^3{C}\cr\cr &+&
10g(\mu^\lambda\partial_\lambda{C})^4{C}+
12g(\mu^\lambda\partial_\lambda{C})^5{C}+\widehat{C}\partial_i\hat{\mu}^{i}+
\hat{C}_i(P^{5i}-\H^{5i})\rangle
\end{eqnarray}

Where we have also included $Q^A$, the contributions of the first
class constraints associated to the antisymmetric field, which
under assumption (\ref{divP}) become a reducible set.

We may now construct the BRST invariant effective action. We
follow the BFV approach {\cite{batalin}} {\cite{mariobfv}}, and we
obtain

\begin{eqnarray}\label{accionefectiva}
S_{eff}&=&\int d^5\sigma d\tau
\left[\mu\dot{C}+\mu_\rho\dot{C}^{\rho}+
P^{\mu\nu}\dot{B}_{\mu\nu}+\Pi_a\dot{X}^a+\hat{\mu}^i\dot{\hat{C}}_i+
\widehat{\mu}\dot{\widehat{C}}+\hat{\delta}(\lambda\mu+\lambda^\rho\mu_\rho+
 \hat{\lambda}_i\hat{\mu}^i+
\hat{\lambda}\hat{\mu})\right.\cr
 &+&\left. \hat{\delta}(\bar{C}\chi+\bar{C}_\rho\chi^\rho+
\bar{\hat{C}}^i\hat{\chi}_i+\bar{\hat{C}}_1\chi_1+\bar{C}_2\chi_2+
\bar{C}_3\chi_3+\bar{C}_4\chi_4)\right]
\end{eqnarray}

Where $\lambda$, $\lambda_\rho$ are the Lagrange multipliers
associated to the diffeomorphisms  constraints while
$\hat{\lambda}_i$ is associated to the constraint on the
antisymmetric field. $\chi$, $\chi^\rho$ and $\hat{\chi}_i$ are
the associated gauge fixing functions. $\chi_1$, $\chi_2$,
$\chi_3$ and $\chi_4$ are the gauge fixing associated to the
reducibility of the  constraints on the antisymmetric field. We
take the following gauge fixing functions:

\begin{eqnarray}\label{gaugecond}
\chi&=&\lambda-\frac{1}{\sqrt{W}} \cr \chi^\rho &=& \lambda^\rho
\cr \hat{\chi}_i&=&\hat{\lambda}_i
\end{eqnarray}

Where we have introduced a non singular metric over the spatial
world volume. This is so, because $\lambda$ transforms as a
density under diffeomorphisms. In some particular cases we may
take $W=1$. The introduction of this metric occurs in a similar
way as for the $D=11$ supermembrane, through the gauge fixing
procedure.

 $\chi_1$, $\chi_2$, $\chi_3$ and $\chi_4$ fix the longitudinal parts,
with respect to the covariant derivative constructed with the
metric that has been introduced, of $\hat{C}^i$,
$\bar{\hat{C}}_i$, $\hat{B}_i$ and $\hat{\lambda}_i$, where
$\hat{\delta}\bar{\hat{C}}_i=\hat{B}_i$, $\hat{\delta}$ denotes
the BRST transformation. We refer to {\cite{mariobfv}} for details of
the construction.

We may now integrate out, or eliminate from the field equations,
the auxiliary field. We obtain finally:

\begin{equation}\label{accionefectiva2}
S_{eff}=\int{d^5\sigma d\tau [\mu\dot{C}+\mu_\rho\dot{C}^{\rho}+
P^{\mu\nu}\dot{B}_{\mu\nu}+\Pi_a\dot{X}^a+\hat{\mu}^i\dot{\hat{C}}_i+
\widehat{\mu}\dot{\widehat{C}}+\frac{1}{\sqrt{W}}\hat{\delta}\mu]}
\end{equation}

where $\hat{\delta}\mu$ is the BRST transformed of $\mu$.

We notice that in the construction

\begin{equation}
\{Q,Q^{\prime}\}_{\cd}=\{Q,Q^{\prime}\}=0
\end{equation}

Where the first bracket is a Dirac bracket, while the second one
is a Poisson bracket. Moreover the Dirac bracket of $Q$ with any
coordinate  of the phase space, excluding $B_{\mu\nu}$, is the
same as its Poisson bracket. However the Dirac bracket of $Q$ with
$\langle P^{\mu\nu}\dot{B}_{\mu\nu}\rangle$ is the same as its
Poisson bracket, in the subspace (\ref{divP}). Consequently the
effective action is BRST invariant, since its kinetic term
transform  as

\begin{equation}
\hat{\delta}S_{eff}=\int{d^5\sigma d\tau
\hat{\delta}[\mu\dot{C}+\mu_\rho\dot{C}^{\rho}+
P^{\mu\nu}\dot{B}_{\mu\nu}+\Pi_a\dot{X}^a+\hat{\mu}^i\dot{\hat{C}}_i+
\widehat{\mu}\dot{\widehat{C}}]}=\int{d\tau\dot{Q}}=0
\end{equation}

provided initial and final conditions on the ghost fields are
imposed {\cite{mariobfv}} {\cite{Teitelboim}}, as usual.

We notice that

\begin{eqnarray}
\langle \frac{1}{\sqrt{W}}\hat{\delta}\mu \rangle &=& \langle
\frac{1}{\sqrt{W}}(\phi+C^{\sigma\lambda}\mu_\sigma\partial_\lambda{C}-
\frac{1}{2}C_1^{\mu\nu,\sigma\lambda}\mu_\sigma\partial_\lambda{C}\mu_\nu\partial_\mu{C}
+ 8g(\mu^\lambda\partial_\lambda{C})^3\cr\cr
&+&10g(\mu^\lambda\partial_\lambda{C})^4 +12g(\mu^\lambda
\partial_\lambda{C})^5 )  \rangle
\end{eqnarray}

is not only manifestly BRST invariant, but also well behaved even
on singular configurations of the induced metric, where its
determinant is zero, in spite of the fact that
$\mu^\lambda=g^{\lambda\sigma}\mu_\sigma$. In fact all these term
may be rewritten in terms of the totally antisymmetric symbol
$\epsilon^{\alpha\beta\mu\nu\lambda}$ and the covariant metric
$g_{\mu\nu}$ only.

\section{Light Cone Gauge Hamiltonian of the  M5-brane}

In the previous section we obtained the general formulation of the BRST
invariant effective action of the M5-brane in the covariant gauge  (\ref{gaugecond}),
we showed that the action may be expressed in a manifestly Lorentz covariant way
(in the target space), without restricting the induced metric to have an inverse.
In this section we will analyze some stability properties of the effective action.
To do so we will start from (\ref{accionefectiva}) and fix  the light cone gauge, which leads to physical Hamiltonian
after elimination of the constraints.

Consider  the light cone gauge fixing conditions:

\begin{eqnarray}
X^{+}&=&\Pi_0^+\tau \\ \Pi^+&=&\Pi_0^+\sqrt{W},
\end{eqnarray}

We  will later on discuss the gauge fixing conditions for the antisymmetric field.

The LCG allows to reduce canonically the phase space to its
transverse part only. This is achieved by solving explicitly the
constraint (\ref{fclass3}) for $\Pi_+$, $X^-$ is eliminated from
the constraint (\ref{fclass4}) provided an integrability condition
is satisfy. This condition is a first class constraint  which
generates the volume preserving diffeomorphisms.

The canonical reduction of the effective action yields after the elimination
of the ghost, antighost fields and Lagrange multipliers to the canonical Lagrangian:

\begin{equation}
\widetilde{L}=\Pi_M\dot{X}^M+P_{\mu\nu}\dot{B}^{\mu\nu}-{\ch}_p
\end{equation}

where

\begin{equation}\label{hp}
{\ch}_p=\frac{1}{2}\Pi^M\Pi_M+g2(y+1)+\Theta_{5i}\Omega^{5i}+\Theta_j\Omega^j
+\Lambda^{\alpha\beta}\Omega_{\alpha\beta}
\end{equation}

$M$ denotes the light cone transverse coordinates $M=1,\cdots,9$.
The explicit expression for $y$ is given in (\ref{y}).
$\Lambda^{\alpha\beta}$ is the antisymmetric Lagrange multiplier
associated to the volume preserving diffeomorphism generated by
$\Omega_{\alpha\beta}$. The explicit expression for
$\Omega_{\alpha\beta}$ is

\begin{equation}
\Omega_{\alpha\beta}=\partial_\beta\{
\frac{1}{\Pi_0^+\sqrt{W}}[\Pi_M\partial_\alpha{X}^M+\frac{1}{4}V_\alpha]\}-\partial_\alpha
\{
\frac{1}{\Pi_0^+\sqrt{W}}[\Pi_M\partial_\beta{X}^M+\frac{1}{4}V_\beta]\}.
\end{equation}

$\Omega_{\alpha\beta}=0$ is the local integrability condition in
order to eliminate $X^-$ from (\ref{fclass4}). There is also a
global integrability condition given by

\begin{equation}\label{intcond}
\oint_c\frac{1}{\Pi_0^+\sqrt{W}}[\Pi_M\partial_\alpha{X}^M+\frac{1}{4}V_\alpha]d\sigma^\alpha=0
\end{equation}

which ensures that $X^-$ is a uniform scalar over the spatial
world volume. $C$ is a basis of homology of  dimension one.
If a compactified target space is assumed, the right
hand side of (\ref{intcond}) may be proportional to an integer.

We have not fix the gauge associated to the antisymmetric field.
There are at least three interesting partial gauge fixing
conditions in this respect:

\begin{itemize}
  \item {The gauge fixing yielding
\[
P^{\mu\nu}=\H^{\mu\nu}
\]
as used in the formulation in {\cite{JS3}}
  }
  \item {The gauge condition

\[
B_{5i}=0
\]
which after double dimensional reduction of the theory yields the
formulation of 4-brane in terms of the antisymmetric field
{\cite{Lamamiaqui}} }
  \item {The gauge
\[
B_{\mu\nu}^T=0
\]
  where $T$ denotes transverse with respect to the derivatives operation.
  It yields after double dimensional reduction to the formulation of the 4-brane
  in terms of the Born-Infeld $U(1)$ vector field
  {\cite{Lamamiaqui}}}.
\end{itemize}

The absolute minimum of the Hamiltonian is obtained at the
configurations satisfying

\begin{eqnarray}
\Pi_a&=&0,\cr g&=&0,\cr
l^{\mu\nu}l^{\alpha\beta}g_{\mu\alpha}g_{\nu\beta}&=&0,
\end{eqnarray}

over any point of the spatial world volume $\Sigma_5$, which we
assume is closed without boundary.

If $l^{\mu\nu}=0$, then the set $\Omega$ in the space of physical
configurations at which the minimum is obtained, becomes the set
of maps $X^a$  from $\Sigma_5$ to the target space, depending on
four linear combinations of the local coordinates. That is, all
maps $X^a$ are functions of at most four of them. It is an
infinite dimensional space of $1,2,3$ and $4$ branes. The
degeneracy of $\Omega$ is analogous to the one that occurs for the
$D=11$ supermembrane. There are string like spikes  in that case,
which are responsible together with supersymmetry for the
continuous spectrum of the supermembrane. The degeneracy of the
world volume may be pictured as lower p-branes emerging form the
world volume which may have free ends on not. It can happen that the
other end is plugged into another disconnected sector of the world
volume. Such configuration is physically equivalent to the
disconnected one, because the tubes not carry any energy. There is
then, instability even in the topology of the membrane
{\cite{nicolai}}. These are general features of all $p$-brane
which are also valid for the 5-brane, in spite of the fact that
the known covariant formulation imposes restrictions to those
singular configurations. The next question that one may ask is if
these instabilities  go away if the antisymmetric field is of
maximum rank, using the fact that it enters in the action as a
quadratic form. That is, if we consider the antisymmetric field
living on a nontrivial higher order bundle (thus avoiding the
$l^{\mu\nu}=0$ configuration), can we still have degenerate
configurations?. The answer to this question is that even if we
assume the antisymmetric field  of  maximum rank at any point of
$\Sigma_5$, there may be string like spikes with zero energy
emerging from the world volume. This is so because the
antisymmetric field $l^{\mu\nu}$ at any point of $\Sigma_5$, has
always at least one zero eigenvalue. Its eigenvector being the
topological vector $V_{\mu}$, which is independent of the induced
metric. In fact

\begin{equation}
l^{\mu\nu}V_\nu=0
\end{equation}

at any point of $\Sigma_5$.

If $V_\mu$ is a zero vector, then $l^{\mu\nu}$ has three
eigenvectors with zero eigenvalue, and the following argument will
be also valid.

We consider the following string like configurations

\begin{eqnarray}
X^{a}&=&X^{a}(Y),\cr \partial_{\mu}Y&=&{\phi}V_{\mu},
\end{eqnarray}

where $\phi$ is scalar field over $\Sigma_5$ which will be
determined.

We may take the family of curves over $\Sigma_5$ tangent, at any
point of an open neighborhood, to $V_{\mu}$. We assume
$V_{\mu}\neq0$ on that open set. We then choose the $\sigma^5$
coordinate along these curves. We then have:

\begin{eqnarray}
\partial_{i}Y&=&0,\cr \partial_{5}Y&=&{\phi}V_{5},
\end{eqnarray}

The solution to this equation is given by

\begin{equation}
\phi=\frac{f(\sigma^5)}{V_5},\quad \partial_{5}Y=f(\sigma^5)
\end{equation}

Where $f$ is an arbitrary scalar field over $\Sigma_5$.

We then conclude, that the quadratic term in the Hamiltonian
arising from the antisymmetric field is zero for any of these
string like configurations. That is, even when $l^{\mu\nu}$ is of
maximum rank at any point of $\Sigma_5$ the world volume can
degenerate to acquire string like spikes with zero energy. We
notice that when $\Pi_a=0$, as we are assuming, the admissible
configurations for the antisymmetric field are restricted by

\begin{equation}
\partial_\mu{X}^{-}=-\frac{1}{4}V_\mu ,
\end{equation}

even so there are admissible configurations with $V_\mu\neq0$ and
of course with $V_\mu=0$. In the latest case we replace in the
above argument $V_\mu$ by one of the eigenvectors of $l^{\mu\nu}$
with zero eigenvalue.

Finally, we discuss topological conditions which prevent
(classically) the existence of degenerated spikes in the world
volume. The condition we are going to discuss is not related to a
BPS bound in the case of the 5-brane.

The integral of the determinant of the induced metric may be expressed as:

\begin{equation}\label{I}
\int_{\Sigma_5}\left(dX^a\wedge{d}X^b\wedge{d}X^c\wedge
{d}X^d\wedge{d}X^e
\right)^\ast\left(\frac{dX^a\wedge\cdots\wedge{d}X^e}{\sqrt{W}}\right)
\end{equation}

Where $\ast$ denotes the Hodge dual. A nontrivial  minimum of this
expression is achieved when

\begin{equation}\label{II}
{}^\ast{U}\equiv*\left(dX^a\wedge\cdots\wedge{d}X^b\right)={\mbox{constant}}\cdot{\mbox{integer}}
\end{equation}

for a set of five maps, $\hat{X}^1,\cdots,\hat{X}^5$. $U$ is then
the curvature of a potential 4-form living on a non trivial higher
order bundle {\cite{mariohob}, \cite{MRT}}. In particular (\ref{I}) implies
that $dX^1,\cdots,dX^5$ are closed, non exact, 1-forms.

We may now show that the set $\Omega$ of configurations at which
this local minimum is obtained is finite dimensional. We consider
any other set of maps $X^1,\cdots,X^5$ on the same higher order
bundle. We have

\begin{equation}
dX^1\wedge\cdots\wedge dX^5=U+\delta{U}
\end{equation}

where $\delta{U}$ is exact. We then obtain

\begin{equation}
\int_{\Sigma_5}(U+\delta{U})*(U+\delta{U})=\int_{\Sigma_5}U*U+
\int_{\Sigma_5}\delta{U}*\delta{U}\geq\int_{\Sigma_5}U*U
\end{equation}

\vskip 0.5cm

The equality is obtained when $\delta{U}=0$. The question is then
if we can perform a deformation of $\hat{X}^1,\cdots,\hat{X}^5$ such that $U$
is preserved. This is  always possible since we may consider

\begin{equation}
\delta{X}^5=f(\hat{X}^1,\cdots,\hat{X}^4)
\end{equation}

where $f$ is an arbitrary scalar over $\Sigma_5$. It seems then
that the same infinite dimensional space of configurations
$\Omega$  will still exists. However, under assumption (\ref{II}),
they can be removed by a volume preserving diffeomorphism. In
fact, if $U$ is preserved under the deformation, $\delta{U}=0$,
then the deformed potential must satisfy

\begin{eqnarray}\label{exact}
U&=&d{V}\cr \delta{V}&=&d{\Lambda}
\end{eqnarray}

However,  under volume preserving diffeomorphisms
with group parameter

\begin{equation}
\xi^{\mu}=\epsilon^{\mu\nu\lambda\rho\sigma}\partial_\nu\lambda_{\lambda\rho\sigma}
\end{equation}

we obtain

\begin{equation}
\delta\left(\hat{X}^a{d}\hat{X}^b\wedge{d}\hat{X}^c\wedge{d}\hat{X}^d\wedge{d}\hat{X}^e
\epsilon_{abcde} \right)=d(^\ast{U}\lambda),
\end{equation}

$^\ast{U}$ being constant by (\ref{II}). We may then cancel the
deformation and remove the degeneracy of $\Omega$ arising from
exact forms in (\ref{exact}). We are thus left with the cohomology
classes of closed forms only. The functional space $\Omega$ of
 minimal configurations is then finite dimensional.

\section{Algebra of diffeomorphisms on a 6 dimensional world
volume}\label{algebra}

We will find in this section all possible field theories, realized
in terms of the fields $X^a$ and $B_{\mu\nu}$  which are
invariant under 6 dimensional diffeomorphisms. To do so, we will
find  the general structure of the Hamiltonian constraint by
imposing the closure of the algebra of diffeomorphism. It turns
out that the only polinomic constraint which depends on
$g_{\mu\nu}$ and not on its inverse is the associated to the
5-brane theory. The algebra is the following

\begin{eqnarray}
\{\phi_\rho,\phi^{\prime}_\mu\}&=&\phi_\mu\partial_\rho\delta+\phi^{\prime}_\rho\partial_\mu\delta
-(\partial_\gamma{P}^{\gamma\delta})l^{\alpha\beta}\epsilon_{\mu\delta\rho\alpha\beta}\cdot\delta
\label{algebra1}\\
\{\phi_\rho,\phi^{\prime}\}&=&(\phi+\phi^{\prime})\partial_\rho\delta-
4(\partial_\lambda{P}^{\lambda\alpha}){D}_{\alpha\rho}\cdot\delta\label{algebra2}\\
\{\phi,\phi^{\prime}\}&=&({\cc}^{\rho\sigma}\phi_\rho+
{\cc}^{\prime\rho\sigma}\phi_\rho^{\prime})\partial_\sigma\delta\label{algebra3}
\end{eqnarray}

where the prime denotes evaluation at a point on the world volume
of local coordinates $(\sigma^{\prime}_1,\cdots,
\sigma^{\prime}_5)$. ${\c}^{\rho\sigma}$ and $D^{\alpha\rho}$ are
local functions of the canonical variables. $\phi_\rho$ may be
realized in terms of the  ``topological'' expression
(\ref{fclass4}) (since it is independent of any metric on the
world volume)

\begin{equation}\label{generador-espacial}
\phi_\rho=\Pi_a\partial_\rho X^a+\frac{1}{4}V_\rho
\end{equation}

where
\begin{equation}
V_\rho=\epsilon_{\rho\alpha\beta\gamma\delta}l^{\alpha\beta}l^{\gamma\delta}
\end{equation}

\begin{equation}
l^{\mu\nu}=\frac{1}{2}(P^{\mu\nu}+\H^{\mu\nu})
\end{equation}

We propose for $\phi$ the expression

\begin{equation}
\phi=\frac{1}{2}\Pi^a\Pi_a+W
\end{equation}

$W=gF(y,z)$ where $F$ is to be determined in order
 to  satisfy (\ref{algebra3}). We notice that (\ref{algebra2})
is satisfied by any scalar density of weight 1, in the sense
$\phi=g\cdot({\mbox{scalar field}})$, since $\phi_\rho$ given by
(\ref{generador-espacial}) generates the diffeomorphisms on the
spatial 5 dimensional sector of the world volume.

The general solution for $F(y,z)$ is given by the space of
solutions of the partial differential equation --We refer to
Appendix A for the detail calculations of (\ref{algebra3})--.

\begin{equation}\label{ecuacionF}
2\frac{\partial F}{\partial z}z+\left(\frac{\partial F} {\partial
z}\right)^2z+y\frac{\partial F}{\partial z}\frac{\partial F}
{\partial y}+\left(\frac{\partial F}{\partial y}\right)^2=
2F-2z\frac{\partial F}{\partial z}-2y\frac{\partial F}{\partial
y},\quad F\neq0
\end{equation}

In particular, the following is a solution of (\ref{ecuacionF}):

\begin{equation}
F=2y+\lambda\sqrt{1+y+z}+\frac{\lambda^2}{8}+2
\end{equation}

for any value of $\lambda$.

The final form of the constraint for this solution is then given
by

\begin{equation}
\phi=\frac{1}{2}\Pi^a\Pi_a+g\left[2y+\lambda\sqrt{1+y+z}+\frac{\lambda^2}{8}+2\right]
\end{equation}

and replacing the expression for $y$ and $z$ we have

\begin{equation}\label{vinculo}
\phi=\frac{1}{2}\Pi^a\Pi_a+g\left[
g^{-1}l_{\alpha\beta}l^{\alpha\beta}+\lambda
\sqrt{1+\frac{1}{2}g^{-1}l_{\alpha\beta}l^{\alpha\beta}+
\frac{1}{64}g^{-1}g^{\mu\nu}V_{\mu}V_{\nu}}
+\frac{\lambda^2}{8}+2\right]
\end{equation}

When $\lambda=0$ we obtain the constraint associated to the
Lagrangian (\ref{LS}), which is quadratic on the field
$B_{\mu\nu}$. It is interesting to study which could be the
Lagrangian associated to the more general constraint
(\ref{vinculo}). It turns out to be the Lagrangian (\ref{LS})
with cosmological term $\frac{\lambda}{4}\sqrt{G}$ added.

Indeed, the  canonical analysis of such an action along the lines
presented in  section (\ref{CA}) yield constraint (\ref{vinculo}).
Explicitly we have

\begin{equation}
L=2n\sqrt{gM}-\frac{1}{4}N^\rho
\hat{V}_\rho+\frac{1}{2}\H^{\mu\nu}\partial_0
B_{\mu\nu}+\frac{\lambda}{2}ng^{1/2}
\end{equation}

from which we obtain

\begin{equation}\label{pi2}
\Pi_a=2\frac{g^{1/2}}{n}(-\dot{X}_a+N^{\lambda}\partial_\lambda
X_a)T -\frac{1}{4}\hat{V}^\rho\partial_\rho X_a
\end{equation}

where

\begin{equation}
T=(M^{1/2}+\frac{\lambda}{4})
\end{equation}

and from (\ref{pi2}) --instead of (\ref{pi1})-- we obtain
(\ref{generador-espacial}) and (\ref{vinculo}) in the
gauge $P^{\mu\nu}=H^{\mu\nu}$ .

We notice in (\ref{vinculo}),  that the term
$gg^{\mu\nu}=\frac{1}{4!}\epsilon^{\mu\nu_1\nu_2\nu_3\nu_4}g_{\nu_1\lambda_1}
g_{\nu_2\lambda_2}g_{\nu_3\lambda_3}g_{\nu_4\lambda_4}
\epsilon^{\nu\lambda_1\lambda_2\lambda_3\lambda_4}$, hence it may
be rewritten in terms of the covariant metric $g_{\mu\nu}$. The
expression (\ref{vinculo}), consequently, may be expressed in
terms of $g_{\mu\nu}$, it is not assumed the existence of the inverse.
In order that the scalar $z$ may be
involved in an expression with that property, it must appear as

\begin{equation}
z^{\alpha},\quad \alpha\leq\frac{1}{2}
\end{equation}

The  expression will then be non-polinomic. The most general
polinomic solution to (\ref{ecuacionF}) which does not depend on
the inverse metric is then

\begin{equation}
F=ay+\frac{1}{2}a^2
\end{equation}

for any real number $a\neq0$. However $a$ must be at least $a>0$
in order to have a bounded from below Hamiltonian. It then may be
rescaled to a fix number since the transformations

\begin{equation}
  \begin{cases}
    \Pi\longrightarrow\lambda\pi & l\longrightarrow{l}, \\
     X\longrightarrow\frac{1}{\lambda}X& a\longrightarrow\lambda^6a,\\
     \phi\longrightarrow\lambda^2\phi &.
  \end{cases}
\end{equation}

leave the canonical Lagrangian invariant. The property, of the
realization of the algebra, for the M5-brane of being only
dependent on $g_{\mu\nu}$ and not on its inverse is preserved when
we consider a dimensional reduction  to $5$ dimensional world
volume. If we impose the double dimensional reduction, in the
sense

\begin{eqnarray}
x^5&=&\sigma^5\cr \frac{\partial}{\partial\sigma^5}&=&0
\end{eqnarray}

we then obtain the following realization of the algebra

\begin{eqnarray}
\{\phi_i,\phi^{\prime}_j\}&=&\phi_j\partial_i\delta+\phi^{\prime}_i\partial_j\delta
-(\partial_k{P}^{kl}){\H^{5m}}\epsilon_{ijlm}\delta\\
\{\phi_i,\phi^{\prime}\}&=&(\phi+\phi^{\prime})\partial_i\delta-
4(\partial_l{P}^{lm})l_{mi}\cdot\delta\\
\{\phi,\phi^{\prime}\}&=&({\cc}^{ik}\phi_i+
{\cc}^{\prime{ik}}\phi_i^{\prime})\partial_k\delta
\end{eqnarray}

where,

\begin{eqnarray}
\phi_k&=&\Pi_a\partial_kX^a+\frac{1}{2}\epsilon_{mnlk}P^{mn}\H^{l}\\
\phi&=&\frac{1}{2}\Pi^2+2g+2(\frac{1}{8}P^{ij}P^{kl}g_{ik}g_{jl}+\H^{i}\H^{j}g_{ij})+\frac{1}{32}
(\frac{1}{4}\epsilon_{mnlk}P^{mn}P^{lk})^2 \\
{\cc}^{ik}&=&4(gg^{ik}+\frac{1}{4}P^{ij}P^{kl}g_{jl}+\H^{i}\H^{k})\\
\H^i&=&\frac{1}{6}\epsilon^{ijkl}H_{jkl}
\end{eqnarray}

This theory incorporates the singular configurations of the metric
to the 4-brane theory. It does correspond  to it in the flat limit
{\cite{Lamamiaqui}} and couples consistently the antisymmetric
field with the induced metric. However a direct check  showing the
equivalence of the canonical Lagrangian, arising from these
constraints, and the usual 4-brane one has not be performed. The
Hamiltonian constraint is now quartic in the antisymmetric field.
We have explicitly checked the closure of this algebra with the
structure functions given. It exactly correspond to the
dimensional reduction of the 6 dimensional algebra previously
obtained.

\section{The subspace of solutions with flat induced metric}

In this section we shall show how can we recover the master
canonical action {\cite{Lamamiaqui}} ---with the flat induced
metric--- starting from the canonical action constructed with the
canonical Hamiltonian (\ref{Hcan}), which is quadratic in the
antisymmetric field. To achieve this goal, we will rely on a
particular solution of the field equations

Lets consider the field equations arising from the canonical
Lagrangian

\begin{equation}\label{Lcan}
L=\Pi_a\dot{X}^a+P^{\mu\nu}\dot{B}_{\mu\nu}-\ch
\end{equation}

where

\begin{equation}\label{Hfclass}
\ch=\Lambda\phi+\Lambda^\alpha\phi_\alpha+\Theta_{5i}\Omega^{5i}+\Theta_j\Omega^j
\end{equation}

is the canonical Hamiltonian with the general form of the
Hamiltonian  constraint obtained in section (\ref{algebra}).

By taking variations of (\ref{Lcan}) with respect to $\Pi_a$ and
$X^a$, $\Lambda$ and $\Lambda^\alpha$ we obtain

\begin{eqnarray}
\phi&=&0\label{phi}\\ \phi_\alpha&=&0\label{phia} \\
\dot{X}^a&=&\Lambda\Pi^a+\Lambda^\alpha\partial_{\alpha}X^a
\label{xeq}\\
-\dot{\Pi}_a&=&\Lambda\frac{\delta\phi}{\delta{X}^a}-\partial_a[\Lambda^\alpha\Pi_\alpha]
\label{peq}
\end{eqnarray}

We now consider the subspace of solutions of the field equations
which satisfy initially

\begin{eqnarray}\label{fijacion}
X^0&=&\tau_{0}\cr X^\alpha&=&\sigma^\alpha \quad \alpha=1,\cdots,5\cr
X^a&=&0,\quad a\geq6 \cr\Pi_a&=&0,\quad a\geq6
\end{eqnarray}

We impose gauge conditions in the Lagrange multipliers

\begin{eqnarray}
\Lambda&=&\frac{1}{\Pi^0}\cr \Lambda^\alpha&=&-\Lambda\Pi^\alpha,
\end{eqnarray}

We then obtain from (\ref{phi}), (\ref{phia}), (\ref{xeq}) that
(\ref{fijacion}) are preserved by the evolution equations and $X^0=\tau$. The
explicit expressions for the constraints (\ref{phi}) and
(\ref{phia}) , (\ref{vinculo}) and (\ref{fclass4}) respectively,
provide us the values of $\Pi_0$ and $\Pi_\alpha$ in terms of the
antisymmetric field and their conjugate momenta, specifically from
(\ref{phia}) we can get the values for $\Pi_\alpha$

\begin{equation}
\Pi_\alpha=-\frac{1}{4}V_\alpha
\end{equation}

and replacing this result in (\ref{phi})  we obtain $\Pi^0$

\begin{equation}
-\frac{1}{2}\Pi_0^2 +
\frac{1}{32}V_{\alpha}V^{\alpha}+g(2y+\frac{\lambda^2}{8}+\lambda
M^{1/2}+2)=0
\end{equation}

\begin{equation}\label{pio}
\Pi_0=2\left(M^{1/2}+\frac{\lambda}{4}\right)
\end{equation}

where $M=1+y+z$.

Replacing the expressions for $\Pi_0$ and $\Pi_\alpha$ in the
canonical Lagrangian (\ref{Lcan})  we finally  obtain the canonical
Lagrangian over a flat induced metric,

\begin{equation}\label{Lflat}
L=P^{\mu\nu}\dot{B}_{\mu\nu}-2\left(M^{1/2}+\frac{\lambda}{4}\right)
+ \Theta_{5i}\Omega^{5i}+\Theta_j\Omega^j
\end{equation}

When $\lambda=0$ we recover the master canonical action
{\cite{Lamamiaqui}} for the Perry and Schwarz formulation
{\cite{JS1}}.

\section{Conclusions}

We performed  a complete canonical analysis of the bosonic
M-theory five brane action
corresponding to the partial gauge fixed formulation of the PST
action where the scalar field is fixed as the world volume time.
This canonical formulation is quadratic in the dependence on the
antisymmetric field and it has second class constraints. We
removed the second class constraints by  proposing an extension of the
 canonical action derived from the covariant action of {\cite{Pasti1}}
 and  constructed a  {\it{master canonical action}} with
first class constraints only,  preserving the locality of the field
theory. We then constructed the associated  nilpotent BRST charge,
assuming a world volume with a compact without boundary spatial part,
and its BRST invariant effective theory. The BRST charge is well
defined even for configurations in which the induced metric has
zero determinant at some point or open neighborhood of the world volume.
It does not require the existence of the inverse
of the induced metric. Consequently, the BRST effective action,
which is manifestly Lorentz invariant in the target space and
manifestly BRST invariant in the world volume, has also the same
property. The singular configurations  have then to be
considered as physical one. This implies the existence of
configurations changing the topology of the M5-brane without
changing its energy. We  expect then  that they will have the same
consequences as for the $D=11$ supermembrane with respect to its
spectrum once the supersymmetry is implemented into the theory.

We obtained the physical Hamiltonian of the theory in the LCG from 
the general effective action (\ref{accionefectiva}) and analyzed
its stability properties explicitly. We showed the existence of singular
configurations, where $g=0$, even for maximum rank of the antisymmetric
field. We also constructed global configurations where
singularities are not allowed at any point of the world volume,
and give a geometrical interpretation of them in terms of higher
order bundles.

Finally, by studying the algebra of 6 dimensional diffeomorphisms we found
the most general structure for the Hamiltonian constraint and we identified
the constraint associated with the bosonic five brane action
 upgraded with a cosmological term as a
constraint with a Born-Infeld type term. The M5-brane may be
characterized from this algebraic point of view as been the only
one whose Hamiltonian constraint is polinomial in the antisymmetric
field and is well defined without  assuming the existence of the
inverse of the induced metric.

We have now all elements to start analyzing the spectrum of the
super M5-brane. In particular to look at the massless states of
the theory. It will be important to relate this problem to its
dual one for the $D=11$ super membrane, where the spectrum was
determined only for flat Minkowski target space using an $SU(N)$
regularization. Another aspect that may follow from
our results is the explicit construction of the Seiberg-Witten
map relating the M5-brane to a non-commutative geometries in terms of the BRST cohomology. In this
sense, the simplectic structure we have constructed may be
important.

\begin{center}
{\bf{Acknowledgements}}
\end{center}
We would like to thank I. Bandos, M. Caicedo, L. Quevedo, D. Sorokin and J. Stephany
for interesting discussions. We also would like to thank the theoretical group at
Department of Mathematics, Kings College  for kind hospitality and 
the Research Deanery, Grant G11, at USB for financial support.


\appendix
\section{Appendix}

We will evaluate (\ref{fclass3}) in terms of the functions $F(y,z)$.

We have

\begin{equation}
\{\phi,\phi^\prime\}=\{\frac{1}{2}\pi^a\pi_a+gF,\frac{1}{2}\pi^{\prime
a} \pi^{\prime}_a+g^\prime F^\prime\}
\end{equation}

which by explicit evaluation yields

\begin{eqnarray}
\{\phi,\phi^\prime\}&=&\pi^a\left[\{\pi_a,g^{\prime}\}F^{\prime}+g^{\prime}
\frac{\partial
F^{\prime}}{\partial z^{\prime}}\{\pi_a,z^{\prime}\}
+g^{\prime}\frac{\partial F^{\prime}}{\partial
y^{\prime}}\{\pi_a,y^{\prime}\}\right]\cr\cr &+& g\frac{\partial F}{\partial
z}\{z,z^{\prime}\}g^{\prime}\frac{\partial F^{\prime}}{\partial
z^{\prime}}+ g\frac{\partial F}{\partial
z}\{z,y^{\prime}\}g^{\prime}\frac{\partial F^{\prime}}{\partial
y^{\prime}}+g\frac{\partial F}{\partial
y}\{y,z^{\prime}\}g^{\prime}\frac{\partial F^{\prime}}{\partial
z^{\prime}}\cr &+& g\frac{\partial F}{\partial
y}\{y,y^{\prime}\}g^{\prime}\frac{\partial F^{\prime}}{\partial
y^{\prime}}
\end{eqnarray}

The Piosson brackets of $P^{\mu\nu}$ and $\H^{\mu\nu}$ is given by

\begin{eqnarray}\label{corchetebasico1}
\{P^{\mu\nu},\H^{\prime\hm\hn}\}&=&\frac{1}{2}\epsilon^{\hm\hn\hr\hs\hl}
\{P^{\mu\nu},\partial^\prime_{\hr}B^\prime_{\hl\hs}\}\cr
&=&-\epsilon^{\hm\hn\hr\mu\nu}\partial^\prime_{\hr}\delta(\sigma^\prime-\sigma)
\end{eqnarray}

we then obtain

\begin{eqnarray}\label{}
\{l^{\mu\nu},l^{\prime\hm\hn}\}&=&\frac{1}{4}\left[\{P^{\mu\nu},
\H^{\prime\hm\hn}\}+
\{\H^{\mu\nu},P^{\prime\hm\hn}\}\right]\cr
&=&\frac{1}{2}\epsilon^{\hm\hn\hr\mu\nu}
\partial_{\hr}\delta
\end{eqnarray}

We also get

\begin{eqnarray}\label{}
gg^{\prime}[\{z,y^{\prime}\}+\{y,z^{\prime}\}]&=&\{\frac{1}{2}g_{\mu\alpha}
g_{\nu\beta}l^{\mu\nu}l^{\alpha\beta},
\frac{1}{64}g^{\prime\hat{\mu}\hat{\nu}}V_{\prime\hat{\mu}}
V_{\prime\hat{\nu}}\}
+ \{\frac{1}{64}g^{\mu\nu}V_\mu V_\nu,
\frac{1}{2}g^{\prime}_{\hat{\mu}\hat{\alpha}}g^{\prime}_{\hat{\nu}\hat{\beta}}
l^{\prime\hat{\mu}\hat{\nu}}
l^{\prime\hat{\alpha}\hat{\beta}}\}\cr&=&\frac{1}{32}g_{\mu\alpha}g_{\nu\beta}
l^{\mu\nu}
V^{\prime\hat{\mu}}\{l^{\alpha\beta},V^{\prime}_{\hat{\mu}}\}+\frac{1}{32}
g^{\prime}_{\mu\alpha}
 g^{\prime}_{\nu\beta}l^{\prime\mu\nu}V^{\hat{\mu}}\{V_{\hat{\mu}},
l^{\prime\alpha\beta}\}\cr&=&
 \frac{1}{32}\left[(l_{\alpha\beta}l^{\hat{\alpha}\hat{\beta}}V^{\hat{\mu}}
 \epsilon_{\hat{\mu}\hat{\alpha}\hat{\beta}\hat{\sigma}\hat{\lambda}})+()^{\prime}\right]
 \epsilon^{\alpha\beta\hat{\sigma}\hat{\lambda}\gamma}\partial_\gamma
 \delta
\end{eqnarray}

and

\begin{eqnarray}\label{}
gg^{\prime}\{z,z^{\prime}\}&=&\frac{1}{64}\frac{4}{64}V^{\nu}
V^{\prime\hat{\nu}}
\{V_{\nu},V^{\prime}_{\hat{\nu}}\}\cr\{V_{\nu},V^{\prime}_{\hat{\nu}}\}&=&
4\epsilon_{\alpha\beta\nu\sigma\lambda}\epsilon_{\hat{\alpha}\hat{\beta}
\hat{\nu}\hat{\sigma}\hat{\lambda}}
l^{\sigma\lambda}l^{\prime\hat{\sigma}\hat{\lambda}}\{l^{\alpha\beta},
l^{\prime\hat{\alpha}\hat{\beta}}\}\cr
&=&4\epsilon_{\alpha\beta\nu\sigma\lambda}\epsilon_{\hat{\alpha}\hat{\beta}
\hat{\nu}\hat{\sigma}\hat{\lambda}}
l^{\sigma\lambda}l^{\prime\hat{\sigma}\hat{\lambda}}\frac{1}{2}
\epsilon^{\alpha\beta\hat{\alpha}\hat{\beta}\gamma}
\partial_\gamma\delta\cr
gg^{\prime}\{z,z^{\prime}\}&=&\frac{1}{4}\frac{1}{16}\frac{1}{16}\left[(
\epsilon_{\alpha\beta\nu\sigma\lambda}
V^\nu
l^{\sigma\lambda})(\epsilon_{\hat{\alpha}\hat{\beta}\hat{\nu}\hat{\sigma}
\hat{\lambda}}
V^{\hat{\nu}}l^{\hat{\sigma}\hat{\lambda}})+
()^{\prime}_{\alpha\beta}()^{\prime}_{\hat{\alpha}\hat{\beta}}\right]
\epsilon^{\alpha\beta\hat{\alpha}\hat{\beta}\gamma}\partial_\gamma\delta
\end{eqnarray}

We finally obtain

\begin{equation}
\{W,W^{\prime}\}=I+II+III
\end{equation}

where

\begin{eqnarray}\label{resultadosI}
{\rm I}&=&\left[\frac{\partial F}{\partial y}\frac{\partial
F}{\partial y}\frac{1}{4}gV_{\sigma}g^{\sigma\lambda}+ '
\right]\partial_\lambda\delta
\cr\cr {\rm
II}&=&\left[\frac{\partial F}{\partial z}\frac{\partial
F}{\partial
y}\frac{1}{4}\left(l_{\mu\alpha}l^{\alpha\lambda}V^\mu+\frac{1}{2}
l_{\alpha\beta}l^{\alpha\beta}V^\lambda\right)+ '
\right]\partial_\lambda\delta\cr\cr
{\rm III}&=&\left[\frac{\partial F}{\partial z}\frac{\partial
F}{\partial z}\frac{1}{64}\frac{1}{4}V^{\lambda}V_\mu V^\mu+ '
\right]\partial_\lambda\delta
\end{eqnarray}

We will now evaluated Poisson brackets related to $X^a$ and $\Pi_a$.

We get

\begin{equation}\label{cpl1}
\{\pi^a,g^{\prime}_{\alpha\beta}\}=-2\partial^{\prime}_\alpha\delta(\sigma^{\prime}-\sigma)
\partial^{\prime}_\beta x^{\prime a}
\end{equation}

which may be used to evaluate

\begin{eqnarray}
\frac{1}{2}\{\pi^a\pi_a,W^{\prime}\}&=&\pi^a\left[\{\pi_a,g^{\prime}\}
F^{\prime}+\{\pi_a,F^{\prime}\}g^{\prime}\right]\cr
&=&\pi^a\left[\{\pi_a,g^{\prime}\}F^{\prime}+g^{\prime}\frac{\partial
F^{\prime}}{\partial z^{\prime}}\{\pi_a,z^{\prime}\}
+g^{\prime}\frac{\partial F^{\prime}}{\partial
y^{\prime}}\{\pi_a,y^{\prime}\}\right]
\end{eqnarray}

we obtain after some calculations

\begin{equation}
\frac{1}{2}\{\pi^a\pi_a,W^{\prime}\}={\rm A}+{\rm B}+{\rm C}
\end{equation}

with

\begin{eqnarray}\label{resultadosA}
{\rm A}&=&\left[2Fgg^{\sigma\lambda}\pi^a\partial_{\sigma}x_a+\quad
\prime\quad\right]\partial_\lambda\delta\cr\cr
{\rm B}&=&\left[\frac{\partial F}{\partial
y}g^{\sigma\lambda}l_{\alpha\beta}
l^{\alpha\beta}\pi^a\partial_{\sigma}x_a+ 2\frac{\partial F}{\partial
y}
g_{\mu\alpha}l^{\mu\sigma}l^{\alpha\lambda}\pi^a\partial_{\sigma}x_a+
\quad\prime\quad\right]\partial_{\lambda}\delta\cr\cr
{\rm C}&=&-\frac{1}{32}\left[\frac{\partial F}{\partial z}g^{\sigma\lambda}
g^{\mu\nu}V_\mu
V_\nu \pi^a\partial_{\sigma}x_a +\frac{\partial F}{\partial
z}g^{\mu\sigma} g^{\nu\lambda}V_\mu
V_\nu
\pi^a\partial_{\sigma}x_a+\quad\prime\quad\right]\partial_\lambda\delta
\end{eqnarray}

We now use (\ref{resultadosI}) y (\ref{resultadosA}) to obtain the
closure of the algebra of constraints, that is to restrict $F(y,z)$ in
order to satisfy (\ref{fclass3}). After some calculations we obtain the following
partial differential equation to be satisfied by $F$,

\begin{equation}
2\frac{\partial F}{\partial z}z+\left(\frac{\partial
F} {\partial z}\right)^2z+y\frac{\partial F}{\partial z}\frac{\partial
F} {\partial y}+\left(\frac{\partial F}{\partial y}\right)^2=
2F-2z\frac{\partial F}{\partial z}-2y\frac{\partial F}{\partial y}
\end{equation}


\bibliography{adcar}

\end{document}